\documentclass[acmsmall]{acmart}

\AtBeginDocument{%
  }

\begin{document}

\setcopyright{cc}
\setcctype{by}
\acmJournal{PACMHCI}
\acmYear{2026}\acmVolume{10}\acmNumber{8}\acmArticle{ISS027}
\acmMonth{11}\acmDOI{10.1145/3859339}

\title[From Prefabricated Content to Situated Generation]{From Prefabricated Content to Situated Generation: A Four-Wave Longitudinal Study of AI-Mediated Museum Surfaces and Spaces}

\author{Yibo Meng}
\orcid{0009-0009-0370-5456}
\authornote{Yibo Meng and Xiujun Chen contributed equally to this work.}

\affiliation{%
  \institution{Cornell University}
  \city{New York}
  \country{United States}
}
\email{yibom2@illinois.edu}

\author{Xiujun Chen}
\orcid{0009-0006-5778-0345}
\authornotemark[1]
\affiliation{%
  \institution{Bartlett School of Architecture}
  \department{University College London}
  \city{London}
  \country{United Kingdom}
}
\email{zcfthev@ucl.ac.uk}

\author{Bingyi Liu}
\orcid{0009-0000-4312-153X}
\affiliation{%
  \institution{University of Michigan}
  \city{Ann Arbor}
  \state{Michigan}
  \country{United States}
}
\email{bingyi.ux@gmail.com}

\author{Xiaolan Ding}
\authornote{Corresponding author.}
\orcid{0009-0006-9956-4560}
\affiliation{%
  \institution{North China University of Science and Technology Health Science Center}
  \city{Tangshan}
  \country{China}
}
\email{13838038961@163.com}

\renewcommand{\shortauthors}{Yibo Meng, Xiujun Chen, Bingyi Liu, and Xiaolan Ding}

\begin{abstract}
Generative AI changes a basic condition of interactive surfaces and spaces (ISS): visitor-facing content can be produced during interaction rather than fully authored before it. We report a longitudinal qualitative study with 23 Chinese science museum practitioners, interviewed annually from 2023 to 2026 (retention 23 to 19) and analyzed using reflexive thematic analysis. Participants' accounts moved from anticipated epistemic and operational risks, through backstage experimentation and visitor-facing trials, toward situated governance and renewed questions about the purpose of museum interaction. Following these changes over time shows how generative AI reconfigured the design problems of museum ISS: generated content appeared under institutional authority; inference delay became publicly and bodily experienced; concurrent touch, gesture, and speech created ambiguous multi-user intent; and post-generation editing could leave curators selecting within an AI-defined possibility space. We conceptualize the resulting shift as a movement from prefabricated display to situated generation and formulate four design propositions concerning provenance at the point of display, spatially appropriate fallback, multi-intent negotiation, and curatorial control before generation. These propositions are grounded in practitioner accounts rather than evaluations of implemented solutions.
\end{abstract}



\maketitle
\section{Introduction}

In 2024, a science museum practitioner described a children's installation in which visitors dragged artifact images across a large timeline and received an AI-generated contextual note for each combination (P12, 2024). The interaction could not be reduced to a new way of retrieving prepared media. A visitor's touch now participated in producing what the museum said, at the moment and place in which the institution said it. This seemingly modest change unsettles a practical premise that has long organized museum interactive surfaces and spaces (ISS): curators author and verify content before an exhibition opens, while an installation presents, retrieves, or rearranges that content during a visit.

Generative AI makes that temporal separation unstable. On large touch surfaces, embodied installations, and multimodal interactive spaces, content may be generated in response to a visitor's touch, speech, gaze, gesture, or spatial position. The result appears in a public setting where other visitors can see both the output and the system's delay or failure. Multiple people may act concurrently, leaving the system to infer whose intent should shape a shared response. In a science museum, the generated response also inherits institutional authority even when no curator has reviewed that particular output. Generative AI in ISS is therefore not only a model-integration problem. It changes the relation among authoring, interaction, public space, and accountability.

ISS research has established that surface orientation, bodily engagement, social visibility, and co-located participation actively shape interaction in museums and other public settings~\cite{ciolfi2007designing,hornecker2006learning,hornecker2006grip,horn2012bats,muller2009display,peltonen2008itsmine}. Yet museum ISS studies have predominantly examined visitor behavior, interaction techniques, or individual deployments. We know less about the practitioners who decide what an installation may generate, how it should respond when inference is uncertain, and what counts as an acceptable failure in a knowledge-authoritative space. Human-AI interaction research offers guidance on uncertainty, initiative, and human control~\cite{amershi2019guidelines,yang2020reexamining,horvitz1999principles}, but these principles do not by themselves explain how control should be exercised across large, public, multi-user interfaces whose outputs carry institutional endorsement.

The rapid entry of generative AI into museum practice also makes a longitudinal view necessary. In 2023, P5 stated that AI should not be integrated into museums; in 2025, the same participant described its contribution to spatial and virtual science communication as ``undeniable.'' A study conducted in either year alone would have produced a different account not only of adoption, but of what required design. Following the same panel across four annual waves allowed us to examine how anticipated risk, backstage experimentation, situated deployment frictions, and reflection on curatorial purpose successively made different ISS problems salient. We analyze these accounts as evidence of changing practitioner discourse and self-reported practice, not as direct observation of every system described.

This study is organized around two research questions:

\begin{enumerate}
  \item RQ1: How did practitioners' accounts of generative AI in museum ISS change over time?
  \item RQ2: What ISS design problems became salient as generative AI entered museum practice?
\end{enumerate}

Our analysis identifies a four-phase trajectory in how practitioners located AI in relation to museum ISS: outside the institutional boundary in 2023; entering through backstage work and bounded visitor-facing trials in 2024; becoming subject to situated governance in 2025; and prompting renewed questions about interaction, learning, and curatorial agency in 2026. Resistance and covert use remain important to this account, but they explain how the design space became accessible; they are not the paper's endpoint.

\begin{figure*}[t]
  \centering
  \includegraphics[width=\textwidth]{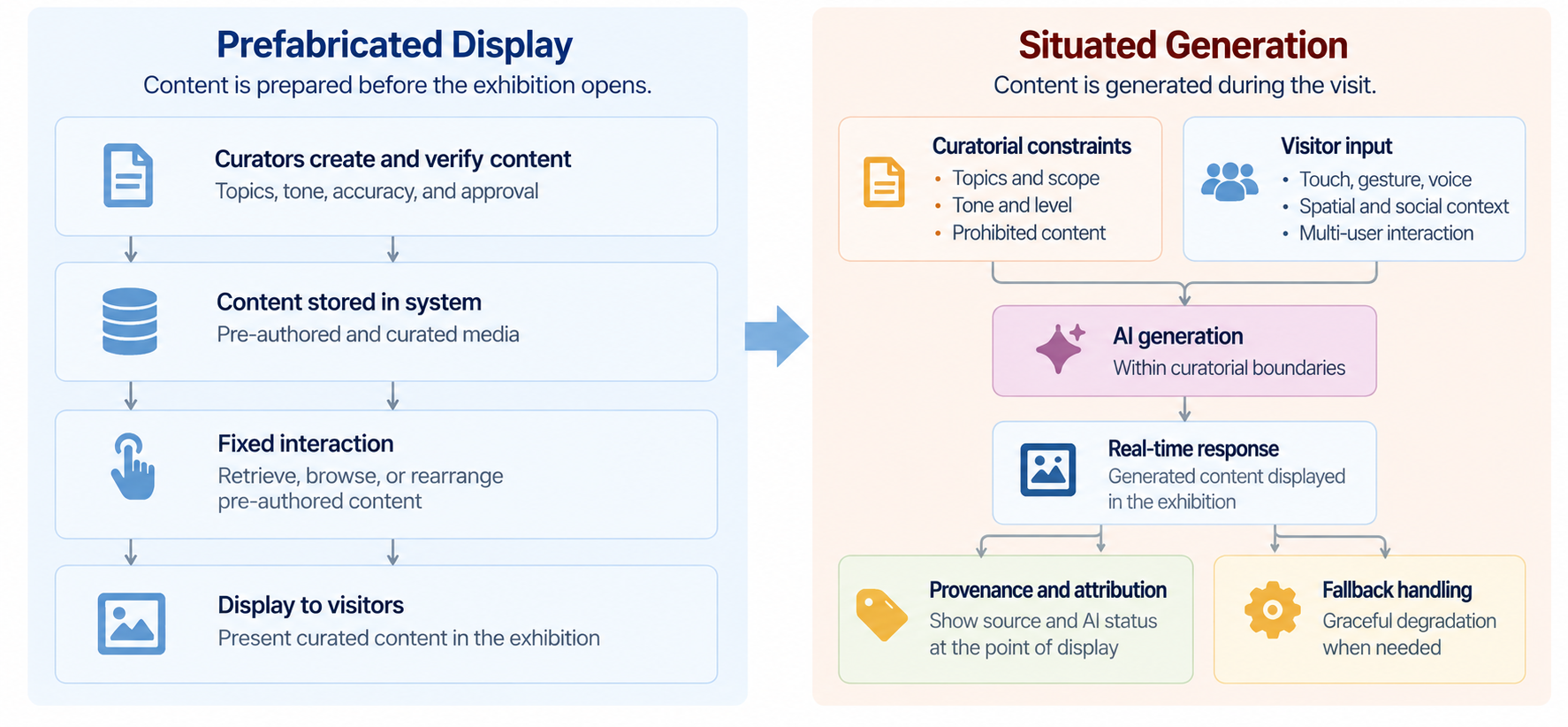}
  \caption{Prefabricated display versus situated generation.}
  \Description{Display versus situated generation.}
  \label{fig:situated-generation}
\end{figure*}

The paper makes three contributions to ISS. First, it shows how generative AI reconfigured the design problems practitioners associated with museum ISS as it moved from anticipated technology, to backstage tool, to a component of visitor-facing interaction. The longitudinal trajectory is the evidence through which this reconfiguration becomes visible, rather than the contribution's endpoint. Second, we identify a shift from \textit{prefabricated display} to \textit{situated generation}: content, presentation, and interaction can become co-produced during a visit, making public latency, multi-user intent, provenance, and failure recovery part of the content-authoring problem. Third, we formulate four empirically grounded design propositions for knowledge-authoritative ISS: making provenance and authorization legible at the point of display; exploring fallbacks that preserve spatial and social interaction rhythms; supporting explicit negotiation of concurrent visitor intent; and enabling curators to specify semantic and epistemic boundaries before generation. Table~\ref{tab:trajectory} summarizes how these problems became salient across the four phases, while Figure~\ref{fig:situated-generation} previews the conceptual shift developed in the Discussion.

\begin{table*}[t]
  \caption{Four-phase trajectory in practitioners' accounts of where AI entered museum ISS practice and which situated design tensions became visible.}
  \label{tab:trajectory}
  \centering
  \small
  \begin{tabular}{p{0.16\textwidth}p{0.24\textwidth}p{0.24\textwidth}p{0.20\textwidth}}
    \toprule
    Phase & Location of AI in ISS practice & Situated tension & ISS design priority \\
    \midrule
    2023: outside the boundary & Anticipated addition to already fragile public installations & Opaque claims, maintenance risk, and institutional accountability & Make provenance, authorization, and failure ownership visible \\
    2024: bounded experiments & Backstage drafting and trials of real-time visitor-facing generation & Public waiting, hallucination, and ambiguous concurrent input & Design fallbacks and multi-intent negotiation for situated use \\
    2025: situated governance & Reported use of local inference, knowledge constraints, and graded review & Oversight needs varied by audience, display location, and consequence & Expose controls that match spatial and epistemic risk \\
    2026: purpose and agency & Richer gesture, multimodal, and adaptive interaction & Attention did not necessarily mean learning; option generation could narrow curatorial judgment & Support semantic agenda-setting before visitor interaction \\
    \bottomrule
  \end{tabular}
\end{table*}

\section{Related Work}

\subsection{Museum ISS as Public and Institutional Interaction}

Museum ISS research shows that interaction is shaped by the spatial, embodied, and social organization of the gallery rather than by an interface alone. Surface orientation, bodily engagement, and the positions visitors take in relation to one another influence participation, interpretation, and learning~\cite{ciolfi2007designing,hornecker2006learning,hornecker2006grip,horn2012bats,falk2005contextual}. Public-display studies likewise show that visibility and co-presence affect whether and how people approach an installation~\cite{muller2009display,peltonen2008itsmine}. Museum technologies can redirect shared attention and produce gaps between designers' intended interpretation and visitors' actual engagement~\cite{grinter2002revisiting,hornecker2008visitor}. Together, these studies establish that museum interaction is not adequately described by an individual user operating a bounded interface: interpretation is organized through bodies, artifacts, bystanders, and the rhythms of a public setting.

A complementary line of work has evaluated how visitors adopt particular museum technologies. An extended technology acceptance model has been used to relate the perceived usefulness, enjoyment, and interactivity of a mobile guide to visitor satisfaction and revisit intention~\cite{kang2018mobile}. A large cross-site study of emerging cultural technologies in Chinese museums mapped visitor engagement with 36 digital systems across 22 sites~\cite{chng2019adoption}. Such studies are valuable for understanding acceptance and use at the visitor--system boundary, but their unit of analysis does not usually include how museum professionals redefine acceptable uses while a technology is being introduced.

Research centered on production and operation makes that institutional work more visible. Schettino followed how front-of-house staff adapted their strategies over a 14-month immersive exhibition~\cite{schettino2016strategies}; Beale and colleagues examined how regional museums and external creative partners developed immersive installations while negotiating skills, resources, and site-specific interpretation~\cite{beale2022digital}. These studies show that an interactive exhibit continues to be learned and remade after installation, and that museum expertise participates in technical design. They also sharpen the gap addressed here. When generative AI composes content during a visit, backstage judgments about sources, interpretation, and failure become conditions of a live public response. The question is therefore not only whether visitors accept an intelligent installation, but how practitioners maintain institutional accountability when authoring extends into runtime.

\subsection{Generative AI in Museum Interaction}

Human-AI interaction research treats probabilistic output, uncertainty, initiative, and human control as design concerns~\cite{amershi2019guidelines,yang2020reexamining,horvitz1999principles}. Mixed-initiative and creativity-support research further asks whether computational systems expand a practitioner's options while preserving meaningful evaluative agency~\cite{lubart2005computers,deterding2017mixed,frich2019mapping}. These concerns acquire different consequences in museum ISS because outputs are public, potentially shared by several visitors, and presented under the authority of a knowledge institution rather than used privately by an individual professional. A delayed or inaccurate response is not encountered only as model behavior: it can interrupt a bodily turn at a surface, become visible to a waiting group, and appear to carry the museum's endorsement.

Recent museum and cultural-heritage systems include recommender systems and ChatGPT-based guides~\cite{trichopoulos2023recommendation,trichopoulos2023crafting}, a context-aware multimodal chatbot~\cite{rachabatuni2024context}, a preference-based guide for a three-dimensional museum tour~\cite{vasic2024llmguide}, and evaluations of LLM chatbots at cultural exhibits~\cite{trichopoulos2026evaluation}. Related work has also examined adaptive visitor presentation and the governance of collections as machine-learning data~\cite{stock2007adaptive,lee2025collections}. This work primarily develops or evaluates visitor-facing guides, collections, or individual systems. It leaves less understood how real-time generation changes design work around public latency, concurrent input, content authorization, and recovery when no practitioner has reviewed the exact response in advance.

\subsection{Professional AI Adoption in Knowledge-Authoritative Domains}

Technology-adoption models commonly explain uptake through perceived usefulness and ease of use~\cite{davis1989perceived,venkatesh2003user}. That framing is useful for predicting acceptance, including visitor acceptance of museum guides, but it captures only part of what is at stake when professionals are accountable for the knowledge a system helps produce. Studies of legal practice first identified skill, mindset, design, and digital-ethics gaps that constrained the provision of technologically enhanced legal services~\cite{janecek2021education}. More recently, interviews with 44 U.S. legal professionals found that lawyers interpreted AI through obligations of competence, confidentiality, candor, and the anticipated effects of AI on legal institutions~\cite{gomezschieber2025attorneys}. In such settings, adoption is inseparable from judgments about professional responsibility.

Education research reveals a related distinction between private assistance and publicly legitimate practice. Barnes and Tour found that ten English-language teachers recognized generative AI's value for preparing teaching materials yet sometimes concealed their use because it could be interpreted as a threat to teaching competence and integrity~\cite{barnes2026dirty}. Fathi's year-long narrative study of eight language teachers similarly describes changing professional identities as teachers integrated generative AI, explicitly challenging a linear progression from resistance to acceptance~\cite{fathi2026becoming}. Across law and education, then, professionals do not simply decide whether a tool is useful. They negotiate where it may enter practice, which obligations remain theirs, and how its use changes the meaning of expertise.

Museum practitioners share this concern with accountable knowledge, but ISS adds a distinctive public and spatial layer. A lawyer or teacher may inspect, revise, or withhold AI-assisted material before it reaches a client or class. In a generative museum installation, an output may appear immediately on a shared surface in response to several visitors' actions. This makes the boundary between backstage experimentation and frontstage use a design boundary: authorization, provenance, timing, and recovery must be expressed through the public interface as well as through organizational policy.

\subsection{Longitudinal Change and Museum Technology in Practice}

Practice-oriented accounts show that technologies acquire meaning and boundaries through situated organizational use~\cite{orlikowski2000using,leonardi2011flexible}. HCI research on non-use similarly treats resistance as a substantive account of professional and social values rather than a deficit to be overcome~\cite{baumer2011implication,satchell2009beyond}. This distinction matters in museums, where credible knowledge practices and institutional identity shape what can legitimately appear on a public interface~\cite{marty2008museum,cameron2008object}. It also suggests that a single measure of acceptance cannot explain why the same practitioner may resist visitor-facing generation while using the technology extensively in private production work.

Longitudinal qualitative research can reveal changes in meaning that a launch study or single interview wave cannot capture~\cite{thomson2003hindsight,rooksby2014personal}. Longer-term HCI evaluations likewise show that initial reactions may not predict sustained use as technologies become routinized and reinterpreted~\cite{klasnja2011evaluating,consolvo2008flowers}. Museum research demonstrates the value of following technology beyond launch. Ioannidis and colleagues traced a substitutional-reality installation from early technical problems through visitor appropriation, pandemic closure, and reopening, arguing for continuing ``design-after-design''~\cite{ioannidis2021wedareyou}. Schettino's 14-month case likewise shows staff practices changing through repeated encounters with visitors and an immersive system~\cite{schettino2016strategies}. These lifecycle accounts follow particular installations; cross-site adoption studies, by contrast, offer broader snapshots~\cite{chng2019adoption}. The present study adds a different longitudinal unit: the same practitioner panel across four annual waves while generative AI moved from an anticipated technology into backstage work and reported visitor-facing trials.

Prior work therefore explains important properties of public ISS, professional adoption, and museum technology-in-use, but it does not yet show how practitioners' changing accounts of generative AI successively redefine what requires design in museum surfaces and spaces. This gap motivates our paired questions: how practitioners' accounts changed over time (RQ1), and which ISS design problems became salient as generative AI entered practice (RQ2). The four phases in the Results preserve that temporal development; the Discussion uses it to identify the shift from prefabricated display to situated generation without proposing a universal stage model of adoption.
 
\section{Method}
 
\subsection{Research Design}
 
This study adopts a longitudinal qualitative research design to examine how science museum practitioners described the changing design, production, and operation of AI-mediated interactive surfaces and spaces (ISS) across four rounds of repeated semi-structured interviews in 2023, 2024, 2025, and 2026. The design was chosen because the ISS issues participants could articulate changed as AI moved from an anticipated addition, to backstage experimentation, to visitor-facing trials and more routinized production. A single cross-sectional study would have captured only one configuration of this design space. The design supports documentation of these trajectories and of how participants accounted for them; it does not allow period effects and direct experience effects to be causally separated, nor does it independently verify the performance of the installations participants described.
 
\subsection{Participants}
 
\subsubsection{Recruitment}
 
Participants were recruited through a combination of purposive sampling and snowball sampling. In the early stage of the study, the research team drew on existing professional networks to issue initial invitations to management and curatorial teams at multiple science museums, thereby establishing a core group of participants. Building on this foundation, participants who had agreed to take part were asked to refer eligible colleagues from their own professional networks, gradually expanding the sample. To ensure adequate heterogeneity in occupational background and age distribution, the research team applied quota controls on occupational category and age group during the snowball sampling process, preventing the sample from becoming concentrated in any single professional group.
 
Inclusion criteria required participants to meet the following conditions: current employment in a Chinese science museum institution in a role related to exhibition design, content planning, digital media development, or collection research, with at least 12 months of experience and plans to continue in this or a related profession; direct observational experience of their institution's ISS practices; and willingness to participate in a longitudinal interview study with a commitment to take part in multiple rounds of follow-up interviews throughout the research period where possible. Participants were excluded if their role was purely administrative with no substantive engagement with exhibition or digital interactive practice.
 
\subsubsection{Recruitment Outcomes}
\begin{table*}[t]
  \caption{Participant Demographics and Attendance by Year}
  \label{tab:participants}
  \begin{tabular}{llllllllll}
    \toprule
    ID & Age & Gender & Occupation & Education & 2023 & 2024 & 2025 & 2026 \\
    \midrule
    P1  & 44 & M & Collection Researcher      & Bachelor & Yes & Yes & No  & No  \\
    P2  & 36 & F & Curator                    & Bachelor & Yes & Yes & Yes & Yes \\
    P3  & 37 & F & Science Communication Researcher & Bachelor & Yes & Yes & Yes & Yes \\
    P4  & 45 & M & Curator                    & Master   & Yes & Yes & Yes & Yes \\
    P5  & 50 & F & Museum Director            & Master   & Yes & Yes & Yes & Yes \\
    P6  & 42 & F & Science Communication Researcher & Master   & Yes & Yes & Yes & No  \\
    P7  & 23 & M & Digital Media Researcher   & Bachelor & Yes & Yes & Yes & Yes \\
    P8  & 27 & F & Digital Media Researcher   & Bachelor & Yes & Yes & Yes & Yes \\
    P9  & 30 & M & Curator                    & Bachelor & Yes & Yes & Yes & Yes \\
    P10 & 31 & M & Curator                    & Bachelor & Yes & Yes & Yes & Yes \\
    P11 & 26 & F & Science Communication Researcher & Master   & Yes & Yes & Yes & Yes \\
    P12 & 22 & M & Science Communication Researcher & Bachelor & Yes & Yes & Yes & Yes \\
    P13 & 35 & M & Museum Director            & Master   & Yes & Yes & Yes & Yes \\
    P14 & 39 & F & Museum Director            & Master   & Yes & Yes & Yes & Yes \\
    P15 & 44 & M & Curator                    & Bachelor & Yes & Yes & Yes & Yes \\
    P16 & 28 & M & Curator                    & Bachelor & Yes & Yes & Yes & Yes \\
    P17 & 29 & F & Digital Media Researcher   & Master   & Yes & Yes & Yes & Yes \\
    P18 & 36 & M & Curator                    & Master   & Yes & Yes & Yes & Yes \\
    P19 & 30 & M & Curator                    & Bachelor & Yes & Yes & Yes & Yes \\
    P20 & 31 & F & Science Communication Researcher & Bachelor & Yes & Yes & Yes & Yes \\
    P21 & 52 & M & Museum Director            & Bachelor & Yes & Yes & Yes & Yes \\
    P22 & 45 & M & Collection Researcher      & Bachelor & Yes & No  & No  & No  \\
    P23 & 34 & F & Digital Media Researcher   & Master   & Yes & Yes & Yes & No  \\
    \bottomrule
  \end{tabular}
\end{table*}

As shown in Table~\ref{tab:participants}, this study recruited 23 participants, all of whom were practitioners currently working in the Chinese science museum sector. Participants ranged in age from 22 to 52 years. The sample comprised 13 male and 10 female participants. Occupational backgrounds spanned five categories: 8 curators, 5 science communication researchers, 4 digital media researchers, 4 museum directors with direct curatorial experience, and 2 collection researchers. In terms of educational attainment, 14 participants held a bachelor's degree and 9 held a master's degree. Participants were drawn from multiple science museum institutions; to reduce the risk of deductive identification in a small professional field, institution names, exact locations, and institution-level counts are not reported in this manuscript. The sample included practitioners from both provincial-level and municipal-level institutions. Regarding attendance, all 23 participants took part in the 2023 interviews; 22 participated in 2024 (P22 withdrew); 21 participated in 2025 (P1 and P22 had withdrawn); and 19 participated in 2026 (P1, P6, P22, and P23 had withdrawn). P1, P22, and P23 withdrew because career transitions had taken them outside the field, while P6 withdrew due to a personal family circumstance. For participants who withdrew, all interview data already collected were retained and incorporated into the analysis for the corresponding years; no replacement recruitment was conducted after withdrawal, in order to maintain the longitudinal continuity of the sample.
 
\subsubsection{Ethics}
 
This study received approval from the Institutional Review Board of the authors' organization. All participants received a detailed informed consent briefing before any interview was conducted, covering the purpose of the study, how data would be used, confidentiality measures, and the right to withdraw. Participants were explicitly informed that their participation was entirely voluntary, that they could withdraw at any point without consequence, and that their institution would not be notified of any withdrawal decision. Interview recordings and transcripts were stored in anonymized form, with participants identified by code (P1 to P23) rather than by name. All information that could lead to identification, including institution names, geographic details, and specific job titles, was removed or obscured during data processing. Raw data were accessible only to core members of the research team and were stored on password-protected encrypted devices, to be destroyed within the specified retention period following the conclusion of the study. Each participant received approximately four US dollars in compensation following each interview; because participants were employed professionals interviewed in relation to their work expertise, this payment was treated as a token of appreciation rather than as primary compensation for labor.
 
\subsection{Data Collection}

Data were collected through one-to-one semi-structured interviews in 2023, 2024, 2025, and 2026. Each interview lasted approximately 55 to 70 minutes, was conducted in Mandarin by the same trained interviewer, and took place either at the participant's institution or by video call when distance or scheduling made an in-person meeting impractical. The same guide and follow-up strategy were used across formats and waves.

The guide contained eight core questions and a closing reflection prompt. It covered participants' professional backgrounds; the operation and content production of ISS installations; attitudes toward and examples of AI use; difficulties encountered; team or institutional disagreement; and the perceived purpose of science communication spaces. Open probes followed participants' responses rather than a fixed sequence. The interviewer maintained a neutral stance and followed mentions of particular tools by asking about context, experience, and reasoning. The full guide appears in the Appendix.

With consent, interviews were audio-recorded, transcribed verbatim within two weeks, checked against the recordings, and anonymized using participant codes. Analysis was conducted on the Mandarin transcripts. Quotations were translated after analysis and checked against the source text for semantic equivalence. Recordings and transcripts were stored on encrypted, password-protected devices accessible only to the core research team.
 
\subsection{Data Analysis}

We used reflexive thematic analysis~\cite{braun2006using,braun2019reflecting}. The primary researcher first read the full corpus repeatedly, then conducted data-driven, line-by-line coding of meaningful segments. Segments could receive multiple codes, and analytic memos recorded coding decisions, emerging associations, and the researcher's developing interpretations. Codes were grouped into candidate themes, which were iteratively reviewed against the Mandarin transcripts, split or merged where necessary, and defined in relation to the research questions.

\begin{figure*}[t]
  \centering
  \includegraphics[width=\textwidth]{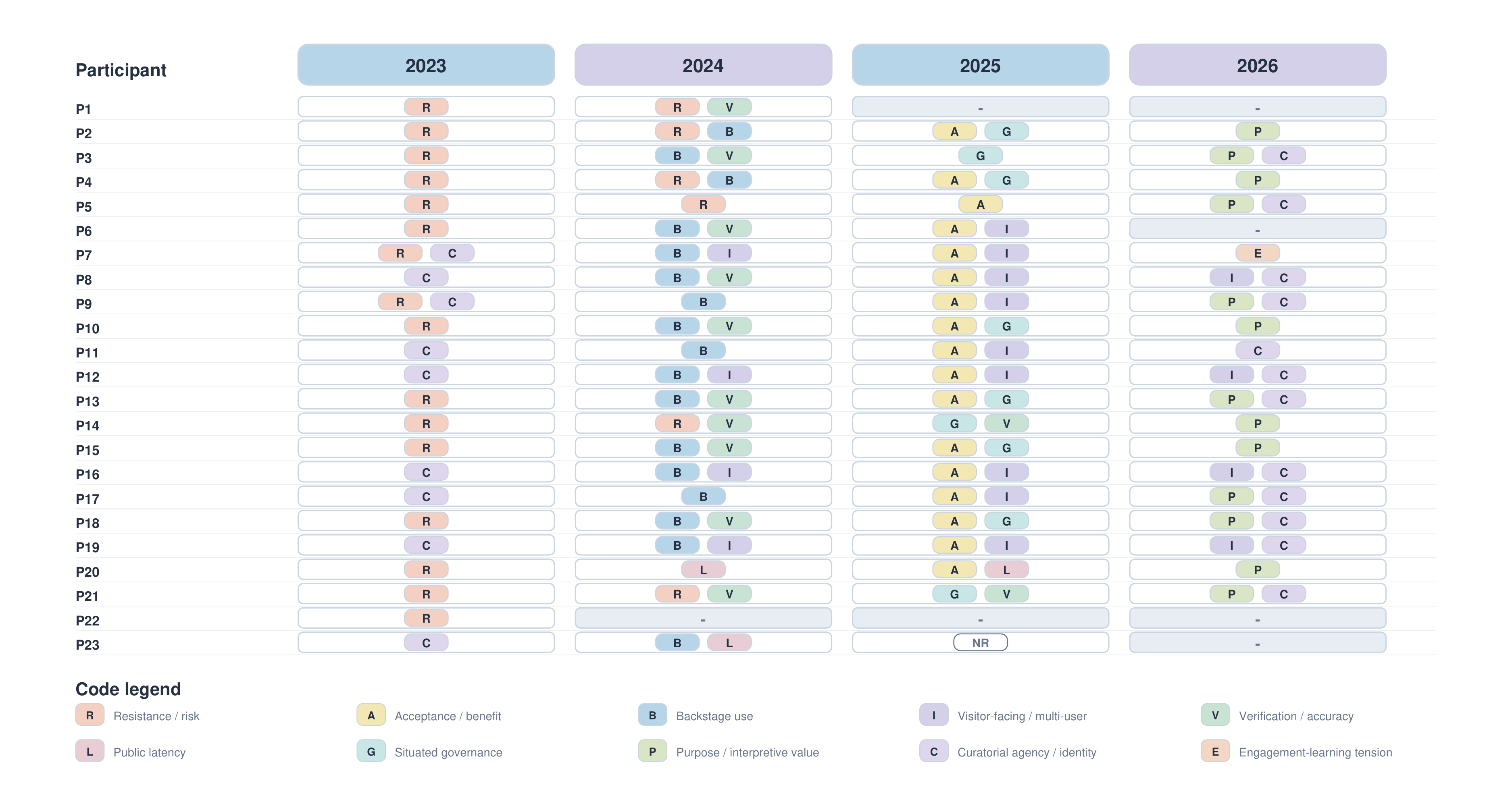}
  \caption{Participant-by-wave coding matrix. Gray denotes no interview; NR denotes no assigned code.}
  \Description{Participant-by-wave coding matrix.}
  \label{fig:participant-wave-matrix}
\end{figure*}

To examine change across waves, we constructed a participant-by-wave trajectory matrix comparing how each participant described AI use, ISS risk, institutional authority, and curatorial value. Cells permitted multiple coexisting codes rather than assigning a participant to one stage. Analysis moved between individual trajectories and group-level patterns, helping us distinguish recurring changes in the continuing panel from apparent shifts affected by attrition or panel composition. Figure~\ref{fig:participant-wave-matrix} presents the completed participant-level summary. Its codes identify salient themes within each participant-wave rather than exhaustive attitudes or mutually exclusive stages.

The four phases are therefore analytic descriptions of the most salient configuration of discourse and self-reported practice in each wave, not mutually exclusive categories or a claim that every participant followed the same sequence. Resistance, curiosity, backstage use, visitor-facing experimentation, and purpose-oriented reflection could coexist within a wave and within one account. Single-participant cases are identified as such rather than treated as prevalence evidence.

Consistent with reflexive thematic analysis, we did not use intercoder reliability as a quality criterion. Quality practices included sustained engagement with the corpus, reflexive memoing, repeated comparison of themes with the source transcripts, and discussion of developing interpretations within the author team. The report retains chronological organization so that changing accounts remain visible while phase-level interpretations are qualified by individual trajectories and within-wave heterogeneity.

\subsection{Researcher Positionality}

The first author's academic background is in human-computer interaction (HCI) and artificial intelligence. This background constituted an epistemological premise that required ongoing scrutiny throughout data collection and analysis. On the beneficial side, training in HCI and AI gave the researcher prior sensitivity to the technical logic of interactive systems, the boundaries of AI capability, and design questions around human-machine collaboration. This helped the researcher understand the technical contexts participants described during interviews and identify detail-level expressions concerning interaction mechanisms and AI performance during coding.

However, this background also carried predispositions that required vigilance. The researcher has long operated within an academic community that maintains an open and broadly positive orientation toward AI technology. This may have inclined the researcher, when encountering participants' expressions of resistance, to interpret these as products of cognitive limitation or insufficient information rather than as professional judgments with their own internal validity. To counter this tendency, the researcher consciously treated participants' resistant discourse as an independent object of analysis throughout, interrogating its internal logic and institutional roots rather than categorizing it simply as resistance to be overcome.

The research team's relationship to the field also mattered. Initial recruitment drew on existing professional networks, which may have increased participants' willingness to disclose sensitive practices such as unsanctioned AI experimentation. This trust strengthened the richness of the data, but it also required analytic caution: participants may have framed their accounts in ways they expected an HCI/AI research team to understand or value. In addition, the researcher's familiarity with ISS as a technical system may have led to over-emphasis on technical problems and breakthroughs during theme construction, with relative inattention to dimensions of institutional culture and professional identity embedded in participants' discourse. To address this, the researcher returned to the data during the review phase specifically to conduct supplementary coding of non-technical expressions, ensuring that the analysis was not systematically occluded by the researcher's disciplinary orientation.

\subsection{Attrition and Panel Effects}

The longitudinal panel declined from 23 participants in 2023 to 19 in 2026. Attrition was not evenly distributed by occupational category. P1 and P22, the only two collection researchers in the original sample, withdrew before the 2025 wave; P1 was also one of the strongly resistant voices quoted in the 2023 analysis. This means that later group-level patterns should not be read as a simple measurement of attitude softening in a stable occupational composition. In the analysis, we therefore treat the four phases as patterns in the available longitudinal corpus and in participants' own accounts of change, while acknowledging that the disappearance of collection-research voices may have contributed to the apparent reduction of resistance in later waves.

The repeated-interview design also introduced the possibility of panel conditioning. Being interviewed annually about AI may have prompted participants to pay more attention to AI, rehearse narratives of change, or present a coherent growth story to the interviewer. The use of the same guide, neutral interview stance, and non-circulation of the guide before interviews mitigated but did not eliminate this risk.

\section{Results}

\begin{figure*}[t]
  \centering
  \includegraphics[width=\textwidth]{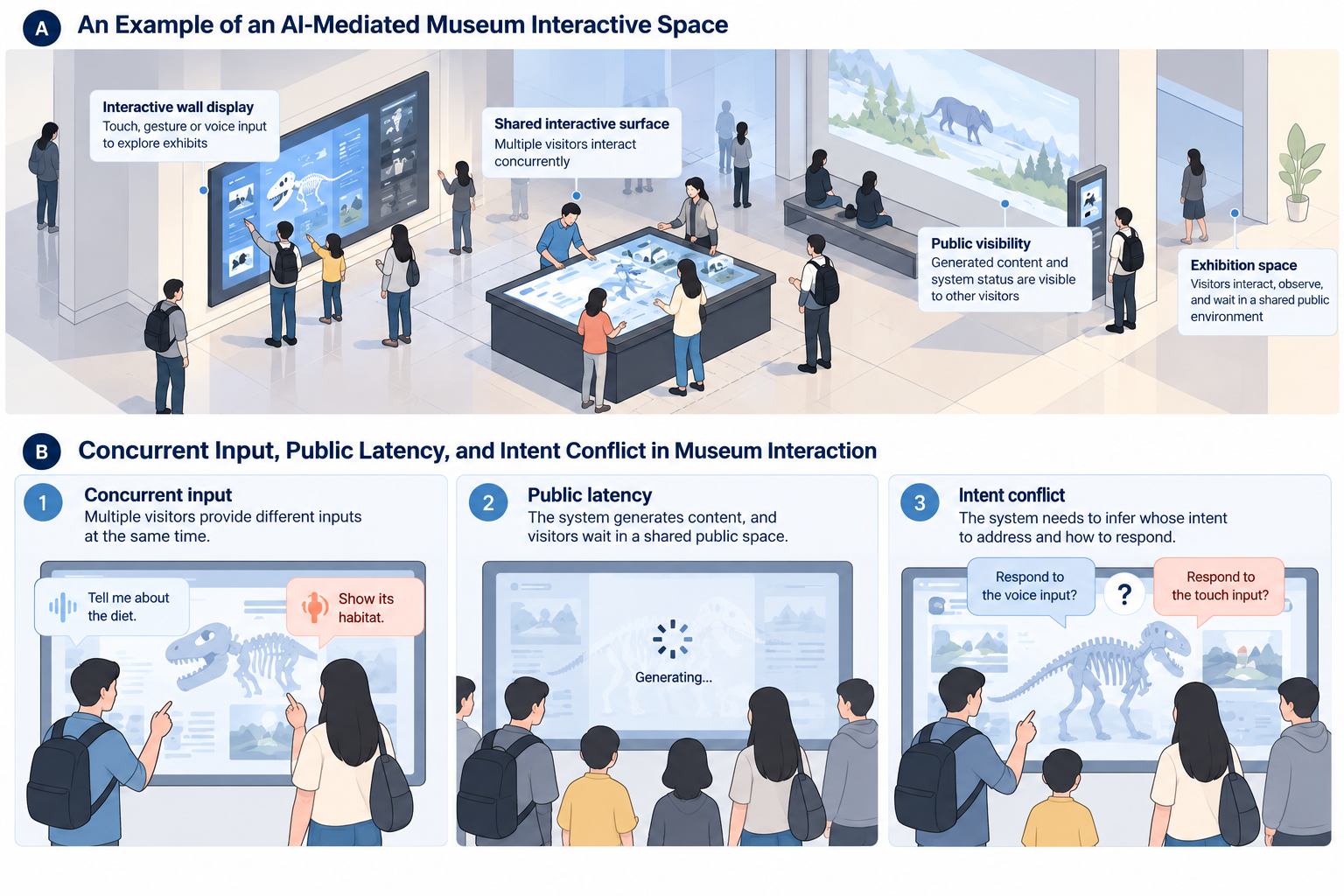}
  \caption{Anonymized museum ISS configurations.}
  \Description{Shared museum ISS configurations.}
  \label{fig:iss-configurations}
\end{figure*}

The following subsections retain interview-wave order so that changes in practitioners' accounts remain visible over time. Across the four phases, the analysis traces both how participants located generative AI in museum practice and which ISS design problems became salient as that location changed. Phase labels name the analytic configuration most visible in a wave; they do not imply discrete calendar-year boundaries, unanimous agreement, or an identical sequence for every participant. The annual design cannot establish when a shift began or whether it persisted between interviews. Single-participant accounts are identified as illustrative cases rather than evidence of corpus-wide prevalence. Figure~\ref{fig:iss-configurations} provides anonymized spatial context for the shared surfaces, public visibility, concurrent input, latency, and intent conflict discussed in the phase accounts.

\subsection{Phase One (2023): Defining AI as Outside the Museum ISS Boundary}

Interview data indicate that in 2023, AI entered the daily discourse of curatorial teams across multiple museums as an explicit topic. However, the prospect of AI mediating interactive surfaces and spaces (ISS) in science communication contexts -- for example, enabling visitors to drive collection retrieval through natural language, children to reorganize exhibition content through gesture, and multi-user intentions to be coordinated and responded to in real time -- provoked publicly articulated resistance across the 2023 corpus. This does not mean that every participant privately held the same position; as shown below, some younger participants simultaneously expressed curiosity or competitive anxiety.

At the level of discourse, this resistance was first articulated as a matter of professional ethics: information presented in educational galleries requires institutional endorsement as a knowledge claim, and its credibility depends on practitioners' specialized training, peer review within the academic community, and the authority that accumulates over time. The introduction of AI was perceived as threatening the very premises of this knowledge production logic.

\begin{quote}
``The label in a museum is not descriptive text. It is a knowledge claim endorsed by the scholarly community. If AI writes it, who is accountable for every sentence? As far as I know, AI has no logic whatsoever. How can something without logic participate in the creation of science communication spaces?'' (P1, 2023)
\end{quote}

In addition, participants commonly drew on concrete experience accumulated from prior ISS operations in science museums: cost overruns in large-screen maintenance, touch recognition failures under high visitor traffic, and system crashes on opening days. Proposals to ``add another layer of AI'' were typically understood, first and foremost, as compounding known risks.

A third dimension of resistance was rooted in the museum's self-understanding as an institutional form. In the professional narratives of several senior practitioners, the distinctive value of the museum lay precisely in its refusal of speed -- in an institutional philosophy that validates meaning through time. Within this framework, the premature adoption of AI was understood as a symbolic loosening of institutional identity, with the threat residing in the signal such a choice would send.

\begin{quote}
``The value of a museum lies in not following trends. It validates meaning through time. If we rush to use AI, we are telling the public that we are no different from everywhere else. There is no hurry. If AI proves genuinely useful later, it will not be too late to adopt it then.'' (P14, 2023)
\end{quote}

These three dimensions -- professional ethics, experience with technological risk, and institutional identity -- appeared in an intertwined form across the 2023 corpus and reinforced publicly articulated resistance to AI integration in educational ISS. Yet this resistance had its limits. A small number of younger participants expressed concern that AI posed a threat to their own professional value -- a reaction that, in itself, implied a partial acknowledgment of AI's capabilities. They identified exhibition captions, guided-tour scripts, catalogue essays, and educational materials as both a substantial part of curatorial labor and visible evidence of professional expertise, while anticipating that generative AI could perform portions of this work.

A small number of younger members also explicitly acknowledged genuine curiosity about generative AI in 2023, but found no safe channel for expressing it. The prevailing discourse of resistance in the museum ISS field had created a form of social pressure that disciplined the space of internal expression.

\begin{quote}
``I really want to try it, but I do not dare say so. It feels like taking the wrong side. This is quite a serious matter. The official position has basically been set, and although I do not think there is anything wrong with that position (resistance to AI in educational science communication), I do feel there is a lot of room for discussion.'' (P11, 2023)
\end{quote}

In producing a surface consensus, the discourse of resistance was simultaneously accumulating suppressed dissent internally. For ISS, this phase established the initial boundary conditions of the design space: generated knowledge claims lacked an accountable author, additional computation threatened already-fragile public installations, and visible adoption risked weakening the museum's institutional identity. The design problem made salient in this phase was therefore not adoption alone, but how a public interactive system could introduce generation without dissolving recognizable responsibility for what the museum said and how reliably it said it.

\subsection{Phase Two (2024): Backstage Experimentation Meets Visitor-Facing Friction}

In the 2024 interviews, AI became visible in two distinct locations of museum ISS practice. Backstage, several younger participants (including P7, P8, P9, P11, and P12) described using generative AI for drafts, audience-specific versions, and planning. At the visitor-facing surface, a smaller set of accounts concerned bounded trials of real-time generation, multimodal input, and shared displays. Other participants continued to articulate the resistant stance of 2023, though some accounts allowed more room for low-stakes assistance. Distinguishing backstage production from live public interaction is important: the former made experimentation institutionally tolerable, while the latter exposed ISS-specific consequences.

\begin{quote}
``My view is basically the same as last year. AI may be able to produce beautiful text and attractive interfaces, but AI should not be used for interactive interfaces or display design in educational contexts. That is a bottom line. But I also acknowledge that perhaps in the future, AI could assist us with other things that are less close to that line.'' (P5, 2024)
\end{quote}

The disagreement in 2024 generated tensions. Because openly declaring support for AI use still carried social costs, early adopters privatized much of their backstage practice. These uses gradually became an ``open secret'': senior colleagues were aware that some curators were using AI, and a tacit equilibrium was maintained even while an explicit declaration could invite sanctions. This low-visibility entry point created room to learn with AI before generated output was placed under the stronger accountability demands of a public installation.

More specifically, early users reported value across several work scenarios. A recurring backstage example was the rapid production of audience-targeted versions of the same material. Participants described exhibition interpretation as requiring separate professional, general-visitor, and child-facing versions, and described generative AI as a starting point that reduced reformatting work while retaining review and judgment.

\begin{quote}
``Writing three versions of the same exhibition guide -- professional, general visitor, and children -- used to be very time-consuming. Now AI can produce a draft and I review and revise it. The time spent on this has genuinely been compressed. It is very convenient with no issues at all.'' (P17, 2024)
\end{quote}

At the level of visitor-facing installation content, participants described a more structural change. They characterized large touch surfaces as having primarily presented or retrieved material prepared before opening, with the range of possible responses constrained by curatorial production time. Their experiments suggested a different configuration in which some exhibition language was generated during interaction.

\begin{quote}
``We tried something new in the children's area: children drag artifact images onto a timeline, and AI instantly generates a brief contextual note explaining what people in that era used the object for. This was impossible before, because the combinations are too numerous to prefabricate. The best interaction we could previously offer was children reaching out and the interface registering their touch in real time -- utterly unengaging. Now AI can handle any question, as long as it falls within the contextual scope of the artifact. The impact on visitor experience is fundamental.'' (P12, 2024)
\end{quote}

Participants also discussed multimodal intent understanding. They contrasted rigid command mappings with systems that attempted to relate natural language to touch, gesture, and visual context. Their accounts framed this inferential combination as promising while acknowledging that interpretation remained inaccurate in some situations.

Among the younger participants who described experimentation in 2024, enthusiasm remained mixed with apprehension. Some attributed this to deference toward senior colleagues and a still-resistant public position, while also reporting that certain backstage uses were tacitly tolerated. More importantly for ISS, their concerns increasingly concerned concrete problems encountered when generation reached an installation.

The first issue was content accuracy. Participants described errors concerning biographies, dates, or historical events as risks to institutional credibility that could not be repaired simply by correcting a later output. Their tolerance for error in public exhibition content was therefore low, while the generative performance they reported in 2024 remained inconsistent.

\begin{quote}
``It described a 1987 installation work as an `early pioneer paying homage to Pop Art,' which is completely untenable academically. It is very convenient during generation, but one must note that all AI-generated content must be reviewed again by a human.'' (P8, 2024)
\end{quote}

The physical performance issues of installations in science communication settings exhibited characteristics specific to exhibition hall contexts. Response latency on large exhibition hall surfaces was public and socially visible, with other visitors present and bodies held in the unnatural posture of waiting. These contextual factors together constituted an experiential framework highly sensitive to response delay, causing latency that was ``acceptable'' in technical testing to become a significant experiential barrier in actual gallery deployment.

\begin{quote}
``The latency problem was more serious than we expected. When visitors touched the screen, AI responses required a wait of three to five seconds. Adults could manage, but children just walked away. The exhibition hall is not home; the experience of standing there waiting is completely different.'' (P20, 2024)
\end{quote}

The additional complexity of multi-user scenarios was also exposed during this period. Participants described shared surfaces on which simultaneous touches left the system without a clear basis for deciding whose action should organize the response.

\begin{quote}
``Misrecognition in multi-touch scenarios is also a problem. When two people touch simultaneously, the system does not know who to respond to. This has not been genuinely solved at the design level. The underlying interaction logic is essentially designed for a single user, and multi-user scenarios are forcibly grafted onto it.'' (P7, 2024)
\end{quote}

\begin{figure*}[t]
  \centering
  \includegraphics[width=\textwidth]{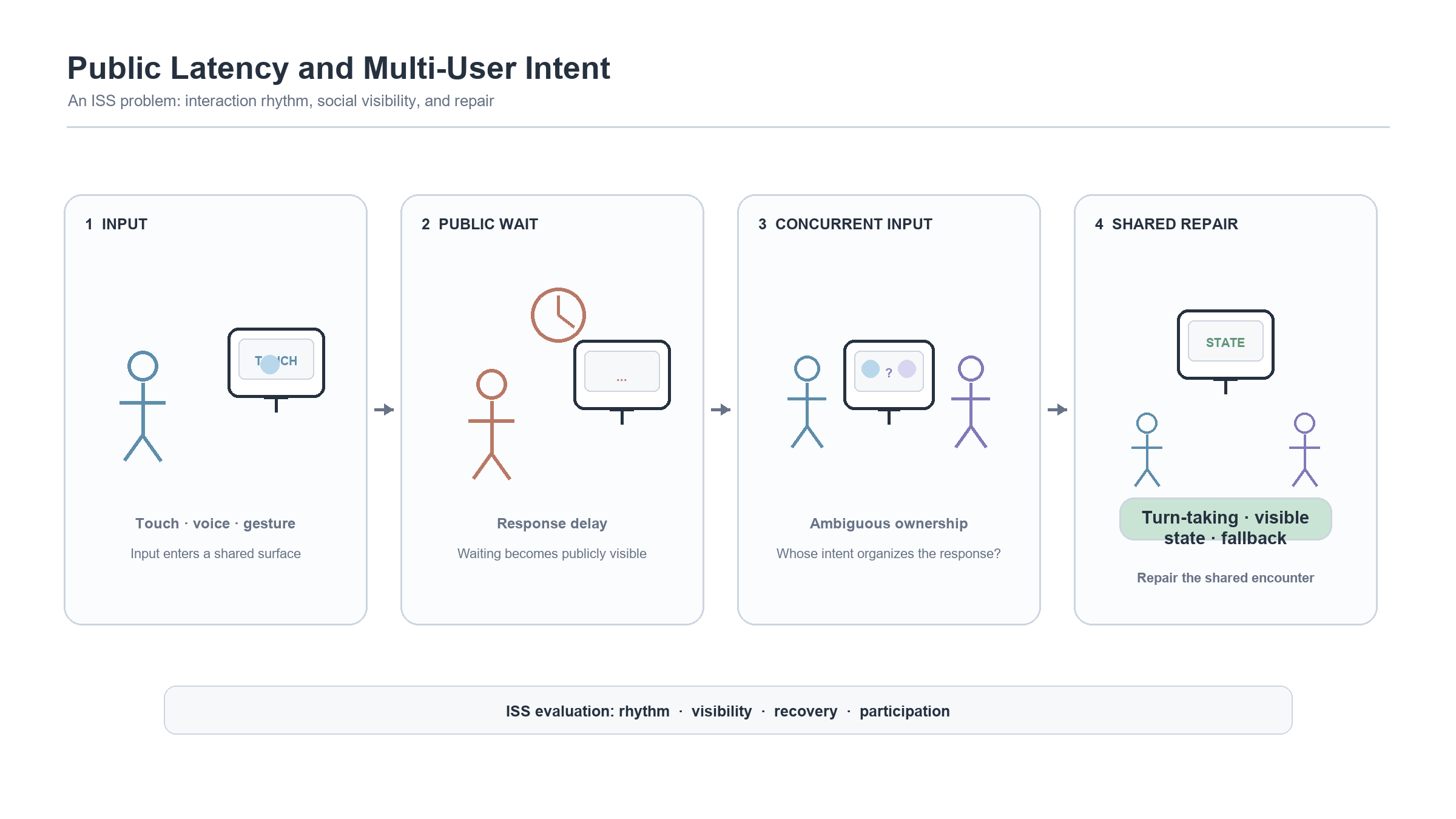}
  \caption{Public latency and multi-user intent in museum ISS.}
  \Description{Public latency and concurrent input.}
  \label{fig:public-latency}
\end{figure*}

In summary, 2024 accounts show AI crossing from backstage production into bounded visitor-facing ISS experiments. That crossing made a set of specifically public and spatial design problems visible: hallucination became an institutionally endorsed claim, inference time became a pause experienced by a body in front of other visitors, and concurrent touch became a coordination problem on a shared surface. Figure~\ref{fig:public-latency} synthesizes these interaction conditions without treating the reported systems as measured deployments. The phase therefore shifts the analytic focus from whether AI was being used to how generation altered the organization of public interaction.

\subsection{Phase Three (2025): Situated Stabilization and Graded Governance}

Entering 2025, resistance was no longer the dominant public register in the continuing panel. Participants more often discussed where an AI function would appear in the gallery, which visitors would encounter it, what input modalities it would interpret, and what review or infrastructure that placement required. AI was increasingly described as a situated ISS component to be governed rather than a technology to be accepted or rejected in the abstract. Even some senior participants explicitly stated they had revised their earlier positions.

The contrast is visible in P5's account. In 2023, P5 rejected AI integration:

\begin{quote}
``No. In anything related to education, science communication, or museums, AI should not be integrated.'' (P5, 2023)
\end{quote}

\begin{quote}
``I think we should be thinking about how AI can be used more effectively -- whether through technical innovation, renovation of interaction mechanisms, or new human factors being taken into account. Overall, I feel that AI's contribution to the forms of museum science communication, particularly spatial and VR formats, is undeniable.'' (P5, 2025)
\end{quote}

P5 provides an illustrative individual trajectory, not a representative path for the continuing panel. We use the comparison to show the kind of within-participant change the longitudinal design can reveal, while the phase interpretation remains grounded in the broader matrix and is qualified by attrition and within-wave heterogeneity.

Participants attributed this shift partly to changes in their reported technical experience. The most concentrated technical complaint in 2024 was latency: participants described visitors touching large screens and then waiting three to five seconds for a response, causing some child visitors to leave entirely. In 2025, participants reported that latency had become less disruptive in the installations they discussed, and several attributed this improvement to cloud-inference upgrades or local inference. They also described greater stability in multi-user touch recognition after targeted redesign, fewer system crashes under peak visitor loads, and partial control of hallucination through domain-specific knowledge base constraints. These are participant accounts of technical change rather than independently verified performance measurements.

\begin{quote}
``That latency problem last year was a real headache. Things are much better now. Children do not stand there waiting anymore -- they keep touching, keep asking, and the entire rhythm of interaction is completely different.'' (P20, 2025)
\end{quote}

Participants also described technologies that had not entered their practical accounts in 2024. In the installations they discussed, body posture, movement trajectory, and spatial position could supplement touch and voice as input for inferring visitor interest. Some participants reported experimental use of facial expression and dwell time to adjust presentation style or difficulty in children's areas. Others described local inference as reducing dependence on unstable gallery networks. These statements document practitioners' perceptions of changing ISS capabilities; the study did not independently inspect the systems or evaluate their sensing accuracy, adaptivity, or reliability.

\begin{quote}
``The system we use now can sense where a child is standing and what direction they are looking -- not just waiting for them to touch something. This is a significant change, because many children simply do not know they are supposed to touch the screen. They just stand there looking. Now the system can actively respond to that `looking.''' (P6, 2025)
\end{quote}

In participants' accounts, improved technical conditions provided more room for strategic maneuver. They described verification processes of varying stringency depending on display location, audience, and the consequence of an incorrect response. Some teams reportedly experimented with adjusting output complexity in children's areas based on perceived age group and engagement rather than fixed versions. Participants working with multilingual content also described AI translation drafts followed by human proofreading as part of their production procedure.

\begin{quote}
``Our use of AI is much more refined than last year. The question is not whether to use it, but at which point to use it, to what degree, and who takes over afterward. None of this existed last year -- we worked it out ourselves, bit by bit.'' (P3, 2025)
\end{quote}

Taken together, the defining characteristic of 2025 was a move toward graded, situated governance. Participants did not describe one level of ``human oversight'' for AI in general. They differentiated oversight by surface location, audience, input modality, public visibility, and the consequence of an incorrect response. The emerging ISS design problem was consequently how to make oversight proportional to the spatial and epistemic stakes of a particular installation. Reported technical stabilization mattered because it enabled participants to articulate this more specific problem; it was not the contribution in itself.

\subsection{Phase Four (2026): Reconsidering Interaction Purpose and Curatorial Agency}

In 2026, a new register of ambivalence about the purpose of museum ISS appeared in some continuing participants' accounts.

As AI use in science communication ISS became more routine in some participants' accounts, a form of reflection emerged that differed from the resistance of 2023: the question in 2023 was ``can AI be used?''; the question in 2026 was ``what should AI be used for?'' P2 marked this new ambivalence, while P14, P17, and P20 each raised the broader question of purpose in different terms. These accounts do not establish that every member of the 2026 panel adopted the same concern, but they show a line of reflection that was not visible in the earlier excerpts.

\begin{quote}
``AI can help us optimize the expressive forms of interactive spaces, and to some extent their expressive content as well. But why do we need interactive spaces at all? That is a question. The market right now is flooded with low-quality AI-generated interactive interfaces. Why should visitors, students, or anyone else pay attention to these interfaces?'' (P14, 2026)
\end{quote}

This inquiry pointed first toward the relationship between form and content. Participants described AI-mediated ISS as visually, behaviorally, and modally richer than earlier installations. P7 reported an internal evaluation in which visitors spent time with a highly mediated installation but recalled less information than visitors reading a conventional label. This is the most concrete displayed account of a possible disconnect between attention and learning; it is a single, unverified case that introduces a design concern rather than establishing a visitor-learning outcome.

\begin{quote}
``We made a very beautiful installation. Children stand there, they can control it with gestures, they can hear sounds, they can see animations. But when we assessed it afterward, the information they could recall about the artifact was less than what they retained from reading an ordinary label. I do not know what we are doing.'' (P7, 2026)
\end{quote}

This was P7's account of an internal evaluation conducted by their institution; the present study did not access the underlying evaluation records or independently compare the two conditions.

This phenomenon touched on the question of ISS's essential function in science communication contexts. The core value of large interactive surfaces in museum science communication settings was understood by participants as the transmission of specific knowledge to specific audiences, in a specific spatial context, through a specific mode of interaction. AI involvement did not automatically answer the questions of ``what knowledge to transmit'' and ``why this knowledge.''

\begin{quote}
``AI can generate very fluent explanations, very beautiful animations, very personalized feedback. But it does not know why this object belongs in this gallery, does not know what this gallery means in this city, does not know where the children who came today are from or what knowledge they carry with them. These are the true context of science communication. AI is not inside that context.'' (P20, 2026)
\end{quote}

This recognition of what participants saw as AI's semantic blindness prompted them to reexamine their own role in ISS. They described earlier work as centered on content prefabrication: deciding what information would appear, in what form, and for which audience. In their accounts, AI could now perform portions of that execution while making an upstream decision more consequential: who decides what this installation should communicate, why, and to whom?

\begin{quote}
``Before, we spent a lot of time thinking about how to write, how to lay things out, how to make text accessible to children. Now AI can do all of that. But I find that I spend more time than ever thinking about `why we are presenting this at all.' Because if I cannot think that through clearly, even the most beautiful AI-generated output has no meaning.'' (P17, 2026)
\end{quote}

However, this renewed affirmation of curatorial judgment faced a counter-question in P11's 2026 account. When AI can already generate an abundance of possible interaction forms, content proposals, and audience strategies, in what sense is the curator's ``judgment'' actually occurring: is it genuine creative decision-making, or is it selection within a space of possibilities that AI has already delimited?

\begin{quote}
``Sometimes I feel that what I am doing is not curating an interactive installation but rather picking from among several options AI has given me what I think is best. These two things feel different, but I cannot articulate where the difference lies.'' (P11, 2026)
\end{quote}

\begin{table}[t]
  \caption{Scope of the Phase Four evidence presented in this paper. The table reports the sources of the analytic claims in the displayed excerpts; it is not a prevalence estimate for the 2026 panel.}
  \label{tab:phase4scope}
  \centering
  \small
  \begin{tabular}{p{0.25\textwidth}p{0.20\textwidth}p{0.39\textwidth}}
    \toprule
    Analytic concern & Sources in displayed excerpts & Evidence boundary \\
    \midrule
    Purpose and interpretive value of ISS & P14, P17, P20 & A broader reflection expressed in distinct terms across three accounts; not a claim about all continuing participants \\
    Interaction richness and learning & P7 & One participant's report of an internal evaluation; records were not available to this study \\
    AI-delimited possibility space & P11 & One phenomenological report of curatorial selection; used to formulate a design proposition, not a prevalence claim \\
    \bottomrule
  \end{tabular}
\end{table}

P11's account is a single phenomenological report, rather than evidence that this concern was shared across the panel. Table~\ref{tab:phase4scope} makes the evidentiary scope of the displayed Phase Four claims explicit. Read alongside the broader purpose-oriented reflections from P14, P17, and P20, P11's account makes a specific ISS design question available for analysis: how might a public interactive system distribute agenda-setting, generation, and judgment among curators, visitors, and AI? We return to this as a bounded design proposition, not as a measured account of a collective 2026 transition.

\section{Discussion}

Read longitudinally, the four phases show practitioners' accounts changing from boundary-making, to differentiated backstage and frontstage use, to situated governance, and finally to questions of purpose and agency. Read as ISS evidence, the same movement made a different design problem salient at each phase: accountable public claims, the embodied and multi-user organization of generation, oversight calibrated to setting and consequence, and curatorial influence over the conditions of generation. The contribution is therefore not a general sequence of AI acceptance. It is an account of how the object of museum ISS design changed as generation moved closer to the public encounter.

This nonlinearity is consistent with longitudinal work in education showing that generative AI can reconfigure professional identity rather than move practitioners smoothly from rejection to acceptance~\cite{fathi2026becoming}. Our account identifies a related process but an ISS-specific consequence: when changing professional judgments are materialized in public surfaces and spaces, professional boundary work becomes design work around visible authorization, embodied delay, shared intent, and repair.

\subsection{Backstage Experimentation and Frontstage Accountability}

The longitudinal account distinguishes two locations of AI in museum ISS practice. Backstage uses produced drafts, audience-specific versions, and planning material that practitioners could inspect before publication. Frontstage uses generated responses during a visitor's encounter with a surface or space. The 2024 pattern of private, low-stakes experimentation matters because it allowed practitioners to develop situated knowledge before exposing outputs to visitors. However, backstage adoption should not be mistaken for evidence that visitor-facing ISS had become institutionally or technically settled.

Moving generation frontstage changed the accountability of the same underlying capability. On a private workstation, an inaccurate draft could be discarded. On a gallery surface, a generated statement became public, socially visible, and readily interpreted as museum-endorsed. Participants' 2023 resistance can therefore be read as a professional account of this transition rather than only as a barrier to adoption. Their concerns joined epistemic provenance, memories of fragile ISS deployments, and the museum's identity as a knowledge institution. This interpretation aligns with HCI accounts that treat resistance and non-use as substantive critiques~\cite{baumer2011implication,satchell2009beyond}, while locating the critique in the public interface itself.

Comparable boundary work appears in other knowledge-authoritative professions. Teachers have reported useful backstage applications while concealing generative AI use under perceived threats to competence and integrity~\cite{barnes2026dirty}; lawyers evaluate AI through professional duties and institutional consequences rather than usefulness alone~\cite{janecek2021education,gomezschieber2025attorneys}. Museum ISS differs in where the boundary becomes operational. Because a generated statement may be immediately visible to visitors, the transition from private assistance to institutionally voiced content must be handled by the interaction design, not only by an upstream policy or an individual professional's later review.

The 2025 movement from general acceptance or resistance toward graded governance also supports practice-oriented accounts in which technology and organizational practice are iteratively constituted~\cite{orlikowski2000using,leonardi2011flexible}. In this study, however, the adjustment was tied to the location and consequence of a public interaction. Participants' accounts therefore make the legibility of authorization a design concern when content crosses from backstage preparation to frontstage display. A visitor-facing system might distinguish prepared, retrieved, and generated material; indicate whether an output has been institutionally reviewed; preserve a stable fallback when generation is uncertain; and give staff a recoverable path from a problematic public response. These are propositions for future interface and deployment work, because the present study did not evaluate their implementation.

\subsection{From Prefabricated Display to Situated Generation}

Much museum ISS design assumes that content is authored and reviewed before a visit, even when visitors can navigate, arrange, or adapt its presentation~\cite{grinter2002revisiting,stock2007adaptive}. Generative AI does not eliminate pre-authored material, but participants' accounts show how it can weaken the temporal separation between authoring and interaction. In P12's example, children arranged artifacts on a timeline and the installation generated contextual annotations for the particular combination. The visitor action was not only a command over existing media; it became part of the conditions under which new exhibition language appeared.

We call this configuration \textit{situated generation}: visitor action, spatial and social context, system inference, and curatorial constraints jointly condition content produced at the time and place of interaction. The concept differs from personalization that selects among prepared assets. It directs attention to generative conditions that ISS designers may need to author: permissible sources, semantic boundaries, uncertainty behavior, session state, and the relation between one visitor's action and what others see. In this configuration, content architecture and interaction architecture become difficult to treat as cleanly separable.

This reframing extends work that treats machine learning as a probabilistic and context-sensitive design material~\cite{dove2017ux,yang2020reexamining}. In a museum ISS, the material is also spatially and institutionally situated. A generated response may need to remain coherent with a nearby artifact, a gallery narrative, the age and position of the current audience, and previous actions on a shared surface. Our accounts suggest that the design object extends beyond individual output quality to the conditions under which an output becomes an appropriate move in an unfolding public interaction.

\subsection{Public Latency and Multi-User Intent}

Participants' accounts also show why model-level performance measures do not fully describe AI performance in ISS. P20 described a three-to-five-second wait after touching a gallery screen: adults waited, while children left. The study did not independently measure that latency, but the account reveals its situated meaning. Waiting occurs through a body held at a surface, in view of other visitors, while the expected rhythm of touch and response breaks down. Prior work has shown that social visibility and the threshold for initiating interaction shape engagement with public displays~\cite{muller2009display,peltonen2008itsmine}; generative delay adds uncertainty after engagement has already begun.

Concurrent input creates a related problem. Multi-user ISS research has developed rich accounts of territoriality, coordination, and collaborative learning around shared surfaces~\cite{scott2003system,dillenbourg2011interactive}. In the installations participants described, AI added inferential ambiguity: speech, touch, gaze, and gesture from several people could plausibly refer to different objects or actions. The design question is not simply whether the system recognizes each signal accurately, but how a shared interface communicates whose intent is active, whether inputs will be merged or serialized, and how participants can repair an incorrect interpretation together.

These accounts motivate an ISS-specific evaluation agenda for future generative systems. Evaluations could examine the spatial experience of waiting, the visibility of system state to bystanders, recovery after failed generation, turn allocation on shared surfaces, and the effect of one visitor's generated content on others' participation. Such measures could complement output accuracy and response time by examining whether generation sustains the social and embodied organization of interaction.

\subsection{Curatorial Agency Before Generation}

By 2026, some participants' accounts had shifted from capability toward purpose. P7 reported that a highly multimodal installation attracted attention without producing the learning outcome their institution expected, although we did not access the underlying evaluation. P11 described feeling that curatorial work had become selection among options whose possibility space AI had already set. These two single-participant accounts are analytically generative rather than prevalence evidence. They caution against treating interaction richness, dwell time, or the number of generated alternatives as proxies for a successful museum ISS without assessing the museum's interpretive purpose.

Mixed-initiative and human-centered AI research argues for explicit allocation of initiative and meaningful human control~\cite{horvitz1999principles,shneiderman2020human}. Third-wave HCI similarly directs attention from instrumental operation toward the meaning created through situated use~\cite{bodker2006second}. The present study makes these concerns more specific for public, generative ISS. Participants' accounts motivate exploring how curators might define why an installation responds, what interpretive relationships it may construct, which claims remain fixed, and how visitor inputs may alter a shared gallery narrative. For outputs produced live in response to visitors, post-generation editing may occur too late to serve as the only form of curatorial control.

\begin{figure*}[t]
  \centering
  \includegraphics[width=\textwidth]{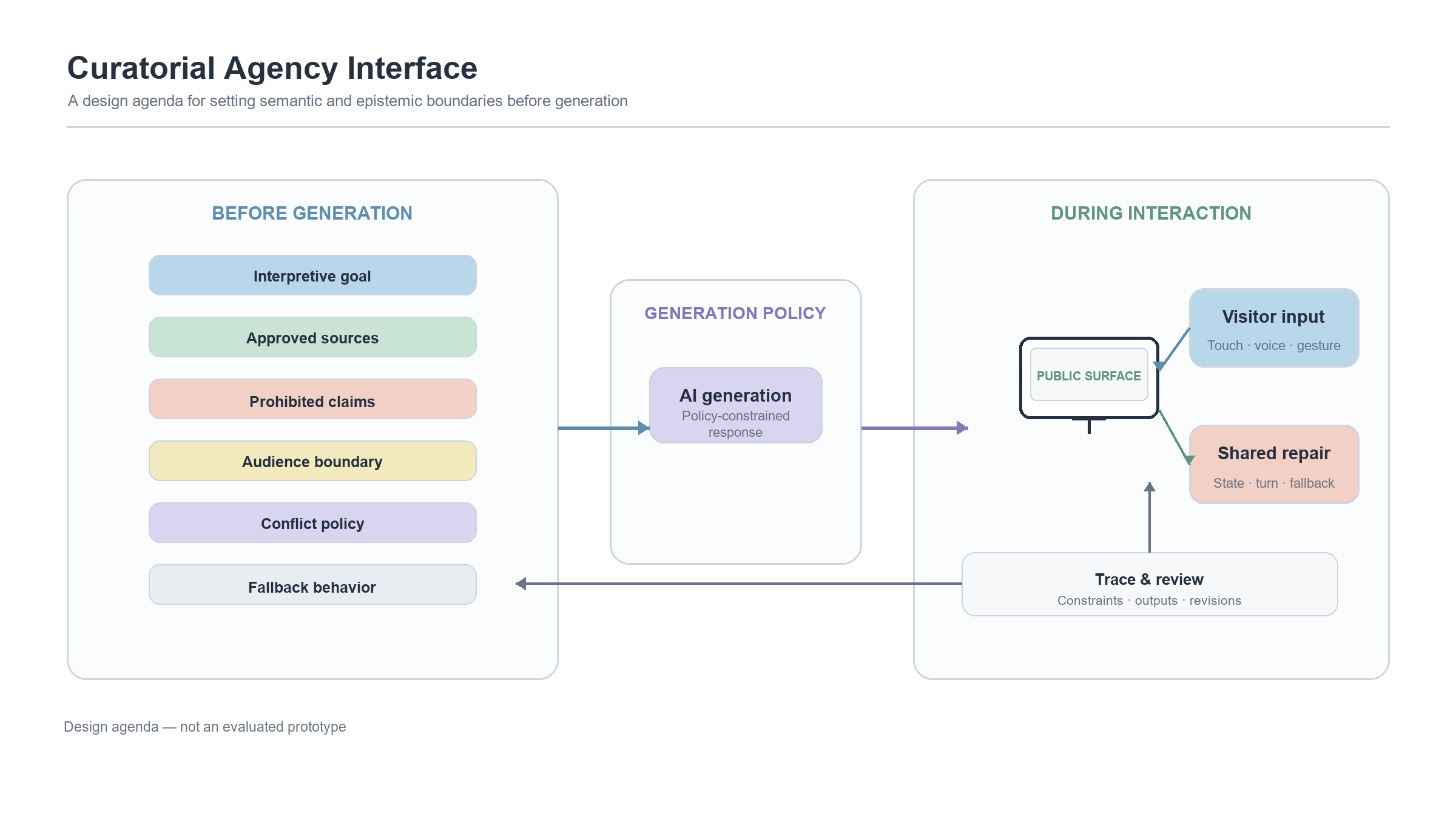}
  \caption{Curatorial control before generation as an unevaluated design proposition.}
  \Description{Curatorial control workflow.}
  \label{fig:curatorial-agency}
\end{figure*}

We use \textit{curatorial agency interface} to name an open design proposition rather than a validated system. Such an interface could let non-technical practitioners establish semantic goals, source and provenance rules, prohibited inferences, audience-specific boundaries, multi-user conflict policies, and fallback behavior before a gallery session. It could also support previewing boundary cases and reviewing how those constraints shaped past outputs. Work on AI explanation, output visualization, complementary team performance, and cognitive forcing functions offers partial resources~\cite{jiang2023graphologue,cai2019effects,bansal2021whole,bucinca2021trust}. Figure~\ref{fig:curatorial-agency} depicts this proposition; future work would need to determine whether it preserves agency under the spatial, social, and institutional conditions of a public encounter.

\section{Limitations}

Several limitations qualify the scope of these findings. First, the design cannot causally separate direct experience effects from broader period effects. The years 2023--2026 were a period of rapid social normalization of generative AI; participants' changing discourse may reflect this broader environment as well as their own reported encounters with AI-mediated ISS. Our evidence for experience-linked change is therefore indirect: participants repeatedly anchored their accounts in specific technical events, workflow experiments, and institutional negotiations, but we do not claim that those events caused the observed trajectory.

Second, the repeated-interview design created risks of panel conditioning and retrospective meaning-making. Participants may have paid more attention to AI because they knew the study would return annually, and later accounts of earlier transitions may have been reconstructed in light of subsequent experience. Capturing four waves mitigated but did not eliminate this risk. Third, attrition changed the occupational composition of the panel: the two collection researchers withdrew before the 2025 wave, which may have contributed to the later reduction of resistant discourse.

Fourth, the study relies on practitioner interviews. We did not collect direct observation, interaction logs, system performance data, visitor learning measures, or the internal evaluation records participants referenced. Technical performance claims in the paper are therefore participant-reported accounts, not independently verified measurements. Fifth, the study is situated in Chinese science museum institutions. Institutional hierarchy, policy environment, and the professional meanings of public endorsement may differ elsewhere. Finally, exact institution counts and locations are not reported in this manuscript to reduce deductive identification risk in a small professional community; this protects participants but limits readers' ability to assess institutional clustering.

\section{Conclusion}

This paper shows how generative AI reconfigured the design problems practitioners associated with museum interactive surfaces and spaces across four annual interview waves. The longitudinal evidence reveals more than changing acceptance: institutional authority became a runtime concern, inference delay became public and embodied, concurrent input became a problem of shared intent, and technical stabilization shifted attention toward situated oversight, interaction purpose, and curatorial agency. We conceptualize these changes as a movement from prefabricated display to situated generation, in which visitor action, spatial context, system inference, and curatorial constraints jointly condition what appears during a public encounter. The resulting propositions concern provenance at the point of display, fallbacks that preserve interaction rhythm, negotiation of concurrent visitor intent, and curatorial influence before generation. They derive from practitioner accounts rather than evaluated interface solutions; future work could test them through direct observation, interaction logs, visitor learning measures, and cross-institutional deployments.

\section*{Ethics and Privacy Statement}

This study involved longitudinal interviews with human participants from 2023 to 2026. The study received institutional ethics approval from the authors' organization. Participation was voluntary, informed consent was obtained, and participants could withdraw without consequence. All names, institution names, geographic details, and other identifying information were removed or obscured. Findings reflect participants' perspectives and do not represent institutional positions. The primary risk was deductive identification through contextual details; this was mitigated through participant coding, institutional anonymization, and restricted access to raw data. Raw recordings and transcripts are stored on encrypted devices accessible only to the core research team and will be destroyed within the specified retention period.

\section*{Data Availability Statement}

The interview transcripts and recordings cannot be publicly shared. The data are longitudinal, institutionally situated, and drawn from a small professional community, making mosaic re-identification possible even after standard anonymization. De-identified excerpts necessary to support the analysis are included in the paper.

\section*{AI Use Disclosure}

Generative AI assistance was used during manuscript preparation for language editing, structural revision, and reference-format checking. The authors reviewed and remain responsible for all analytic claims, interpretations of interview data, citations, and final text.

\begin{acks}
The authors acknowledge the support of their organizations.
\end{acks}

\bibliographystyle{ACM-Reference-Format}
\bibliography{reference}

\appendix

\section{Interview Guide}

The following guide was used consistently across all four annual rounds of interviews (2023 to 2026). The same substantive questions were used in each round to ensure longitudinal comparability, while allowing flexibility for follow-up based on participants' responses. Interviewers did not express positive or negative positions toward AI and did not proactively name specific technology products. If participants mentioned specific tools, interviewers followed up on the context and experience of use without steering evaluative judgments.

\begin{enumerate}
  \item Could you briefly introduce yourself?
  \item What interactive surfaces or interactive space installations does your institution currently have for visitors, and what role do they play in your exhibitions?
  \item What does the process of preparing content for these interactive installations look like?
  \item What is your view on introducing AI into science communication interactive spaces?
  \item Over the past year, have you or your team tried using AI in any part of an interactive installation?
  \item If you or your team have tried using AI in any part of an interactive installation, did you encounter any difficulties or problems in the process?
  \item What is the attitude of your team or institution toward introducing AI into interactive spaces?
  \item Do you think AI involvement has had any effect on what science communication interactive spaces are trying to achieve?
\end{enumerate}

The closing prompt asked: over the past year, was there a particular experience related to AI or interactive installations that made a strong impression on you? Did that experience influence your working approach or thinking afterward? Is there anything important about AI and science communication interactive spaces that we have not yet covered?

\end{document}